\documentclass[aps,prd,groupedaddress,showpacs]{revtex4}

\usepackage[parfill]{parskip}    
\usepackage{graphicx}
\usepackage{amssymb}
\usepackage{epstopdf}
\usepackage{color}
\usepackage{float}
\usepackage{amsmath}
\usepackage{mathtools}

\usepackage{graphicx}
\usepackage{pdfpages}
\usepackage[colorlinks=true,citecolor=blue,urlcolor=magenta,breaklinks]{hyperref}

\begin{document}

\title{Propagation of anisotropic gravitational and electromagnetic waves at very high energies. }

\author{J. Mestra-Páez}
\email{jarvin.mestra@ua.cl}
\affiliation{Departamento de F\'isica, Facultad de Ciencias Básicas, Universidad de Antofagasta, Casilla 170, Antofagasta, Chile.}

\author{Alvaro Restuccia}
\email{alvaro.restuccia@uantof.cl}
\affiliation{Departamento de F\'isica, Facultad de Ciencias Básicas, Universidad de Antofagasta, Casilla 170, Antofagasta, Chile.}

\author{Francisco Tello-Ortiz}
\email{francisco.tello@ua.cl}
\affiliation{Departamento de F\'isica, Facultad de Ciencias Básicas, Universidad de Antofagasta, Casilla 170, Antofagasta, Chile.}

\begin{abstract}
We analyze the dispersion relation for an anisotropic gravity-electromagnetic theory at very high energies. In particular for photons of very high energy. We start by introducing the anisotropic gravity-gauge vector field model. It is invariant under spacelike diffeomorphisms, time parametrization, and $U(1)$ gauge transformations. It includes high-order spacelike derivatives as well as polynomial expressions of the Riemann and field strength tensor fields. It is based on the Ho\v{r}ava-Lifshitz anisotropic proposal. We show its consistency, and the stability of the Minkowski ground state. Finally, we determine the exact zone at which the physical degrees of freedom, i.e. the transverse-traceless tensorial degrees of freedom and the transverse vectorial degrees
of freedom propagate according to a linear wave equation. This is so, in spite of the fact that there exists in the zone  a non-trivial Newtonian background of the same order. The wave equation contains spatial derivatives up to the sixth order, in the lowest order it exactly matches the relativistic wave equation. We then analyze the dispersion relation at very high energies in the context of recent experimental data. The qualitative predictions of the proposed model, concerning the propagation of highly energetic photons, are different from the ones obtained from the modified dispersion relation of the LIV models.  
\end{abstract}
\maketitle

\section{Introduction}
An essential issue in theoretical physics concerns the validity of Lorentz invariance at all energy scales. At low energies compared to Planck energy, the validity seems to be exact, however, one may wonder if that is the case at very high energies. In astrophysical scenarios, one can observe more energetic phenomena (of the order of PeV) than those that one can test experimentally on ground accelerators. Unfortunately, the energies of the detected phenomena are still far away from the Planck scale. Despite this difference in energy levels, in scenarios involving great distances, these effects can accumulate allowing estimates of physical parameters to be made, for example by measuring the time delay between particles of different energies  generated by the same astrophysical source.

Recent estimations of the parameters that characterize the effects of the violation of the Lorentz invariance and the great advance coming from multi-messenger astronomy suggest that in the future there will be remarkable improvements in the characterization of these effects on different sectors, for example, the detection of ultra-high energy photons and the data of the systematic delay time between photons detected by the Fermi-LAT and Fermi-GRB detectors have been used to restrict the values of the parameters that characterize the Lorentz Invariance violation of the photon sector \cite{xu2018regularity,zhang2015lorentz,Ellis2019,Vasileiou2013}. Similar restrictions have been reported for electrons \cite{LiMa2022} from the analysis of gamma rays from the Crab Nebula \cite{Cao2021} and for neutrinos \cite{Zhang2022}
 using data from the Gamma Ray Burst (GRB) catalog (\url{https://user-web.icecube.wisc.edu/~grbweb_public/index.html}) and dates from The IceCube Observatory \cite{AbbasiEtal2021}.  

Fermi-LAT and more recently LHAASO reported the detection of very high and ultra-high energetic photons from several gamma-ray sources \cite{abdollahi2022incremental,Cao2021,CaoEtal2021}. In particular, the detection of more than 530 photons at energies beyond 100 Tev up to 1.4 PeV from different ultra-high  gamma-ray sources. 

Although more experimental data is required to extrapolate a conclusion, it is very unlikely to observe photons with energy higher than 400 TeV if the relativistic principle is maintained. In fact, most extragalactic photons should interact with lower energy photons from the cosmic microwave background CMB and the extragalactic background light EBL, this determines a threshold for relativistic  photons which apparently is violated by the recently detected photons of the order of the PeV. According to estimates which have been given in recent literature the Lorentz violation scale, characterized by the Lorentz violation parameter $E_{\text{LV}}$,  could be  $3.6\times 10^{17}\text{ GeV}$. It has been argued that the violation could be compatible with Superstring theory and M-Theory and also with Loop Quantum Gravity (relativistic theories) due to a spontaneous breakdown of Lorentz symmetry by the ground state of the theory. 

In this paper we consider a different approach and introduce a model inspired by the anisotropic scaling between space and the time proposed by Ho\v{r}ava \cite{Horava2009}, using previous ideas of Lifshitz,  including the interaction terms involving the derivative of the lapse, in a foliated decomposition of the space and time, introduced in \cite{BlasPojolasSibiryakov2010}. They allow us to have a consistent model, free of instabilities, describing anisotropic gravity coupled to a gauge vector field.

Ho\v{r}ava's proposal describes gravity at very high energies in an anisotropic space and time. At very high energies the relativistic symmetry breaks down, and the potential of the Hamiltonian includes then all interactions with high-order spatial derivatives of the Riemann tensor, up to $z=3$ terms (sixth order in spatial derivatives), which implies a dimensionless coupling constant of the action. This geometric structure ensures in principle the renormalizability by power counting of the theory, which in fact has been proven.

 If on the same ground, one would like to introduce other fundamental forces in nature, for example, the electromagnetic force, taking into account that space and time are now described by a foliation of spacelike manifolds parametrized by the time coordinate, one is naturally led to consider anisotropic electromagnetic interactions as well as gravitational ones. Under this assumption, one can start with an anisotropic Ho\v{r}ava action in $4+1$ dimensions, for the electromagnetic interaction, and then introduce a dimensional reduction to $3+1$ dimensions as proposed in \cite{BellorinRestucciaTello2018b,RestucciaTello2020,Restuccia-Tello_2020}  or, as will be proposed here, directly introduce in $3+1$ dimensions all higher order spatial derivatives of the Riemann tensor and the electromagnetic field strength. In both cases the coupling must be determined to fit the experimental data, taking into account that quantum electrodynamics is a very well-established theory. In this sense, the new model has fewer restrictions between its couplings parameters compared to the former one. In any case, the relevant modifications to the relativistic theory should occur at very high energies, on the order of $10^{17} \text{ GeV}$.

Geometrically, the model we propose is formulated on a $3+1$ foliated manifold, it is invariant under spacelike diffeomorphisms that preserve the foliation of the manifold, time reparametrizations, and under $U(1)$ gauge transformations acting on the gauge vector field describing electrodynamics.

This model predicts modified dispersion relations (MDR) for both the gravitational and the electromagnetic sectors with free coupling constants whose value should be restricted by conditions of consistency of the theory and existing experimental data as well as the data to be obtained in the following years from multimessenger astronomy.
This MDR has  different qualitative consequences, with respect to the attenuation of highly energetic photons interacting with background photons, than the ones arising from the LIV approach.

Models that predict modified dispersion relations have been proposed in the context of String theories and D-Branes in \cite{LiMa2021A} and on Loop Quantum Gravity in \cite{Li2022speed}. In the context of anisotropic gravity, as mentioned above, a model was obtained by performing a dimensional reduction of a Ho\v{r}ava-Lifshitz anisotropic gravity model in $4+1$ dimensions. In that case, it was necessary to include up to $z = 4$ spacelike derivative terms in the potential to have a power counting renormalizable theory in 4+1 dimensions. The resulting 3+1 model has coupling constants for gravity and electromagnetism, not all independent \cite{BellorinRestucciaTello2018b,RestucciaTello2020}. We take in this paper a different approach by leaving free the coupling constants with the possibility that they can be adjusted subsequently as the experimental data emerges. This approach was also proposed in a different context in \cite{PospelovShang2012,daSilva2011alternative,kimpton2013matter}.

In section 2 we introduce the anisotropic model and obtain its Hamiltonian structure. The constrained system is completely consistent. To obtain a power counting renormalizable model we consider up to $z=3$ spacelike terms in the potential. In section 3 we analyze the stability of the Minkowskian background and the existence of a wave zone, it requires a set of restrictions on the coupling constants. In section 4 we obtain the evolution equations for the physical degrees of freedom in the wave zone of the theory, i.e. the transverse-traceless tensorial modes and the transverse vectorial ones. These are wave equations with high spacelike derivative terms. We discuss the dispersion formula in the context of the known experimental data. In section 5 we give the conclusions of the work.

\section{Anisotropic gravity-gauge field coupling}

In this section, we introduce the model describing the pure anisotropic gravity-gauge vector field coupling. Once this model and its properties \i.e., symmetries and constraints structure are presented, we check the stability of the reduced Hamiltonian. This analysis is relevant since the Hamiltonian must be positively defined in order to avoid exponential instabilities (ghost fields), rendering the theory to be a well-defined one.  

\subsection{The model}

The action of the model we propose is 
\begin{equation}\label{eq01}
    S=S_{\text{H-L}}+S_{\text{EM}},
\end{equation}
where $S_{\text{H-L}}$ is the Ho\v{r}ava-Lifshitz action
\begin{equation}\label{eq1}
    S_{\text{H-L}}=\int dt\int_{\Sigma_{t}} d^{3}x N \sqrt{g}\left[K^{ij}K_{ij}-\lambda K^{2}+\mathcal{V}_{\text{g}}\right] -\beta\oint_{\partial\Sigma_{t}} dS_{i}\left(\partial_{j}g_{ij}-\partial_{i}g_{jj}\right),
\end{equation}
where $\mathcal{V}_\text{g}$ is the most general scalar constructed from the spacelike derivatives of the Riemann tensor of the leaves of the foliation and $a_{i}\equiv \partial_{i}\ln N$. The potential contains at most six spacelike derivatives in order to have an overall dimensionless coupling constant, which we take to be 1. And $S_{\text{EM}}$ is the following electromagnetic action
\begin{equation}
	S_{\text{EM}}= \int dt\int_{\Sigma_{t}}d^{3}x N\sqrt{g}\left[\frac{1}{2} \left(\frac{F_{0i}}{N}+\frac{N^{k}F_{ki}}{N}\right)\left(\frac{F_{0}^{\:\:i}}{N}+\frac{N^{k}F_{k}^{\:\:i}}{N}\right)-\mathcal{V}_{\text{EM}}(A_{i},g_{ij})\right]\nonumber ,
\end{equation}
where the potential $\mathcal{V}_{\text{EM}}$ includes all scalars with spacelike derivatives up to sixth order, constructed from the contraction of the field strength $F_{ij}$ with the Riemann tensor, $a_{i}$ and itself. The first term in the bracket is the Lagrangian density for the electromagnetic interaction $-\frac{1}{4} F_{\mu\nu}F^{\mu\nu}$, without the term $-\frac{1}{4}F_{ij}F^{ij}$ which is contained in the potential $\mathcal{V}_{\text{EM}}$, expressed in terms of the ADM metric for the foliated manifold.

We will denote $\pi^{ij}$ and $E^{i}$ the conjugate momenta of $g_{ij}$ and $A_{i}$ respectively.

The Hamiltonian density describing the pure anisotropic gravity-gauge field coupling at the KC point is given by 

\begin{eqnarray}
	\label{Hamiltoniandensity}
	\mathcal{H}=N\sqrt{g}  \left[ 
	\frac{\pi^{ij}\pi_{ij}}{g}+\frac{E^{i}E_{i}}{2g}+\mathcal{V}(g_{ij},N,A_{i}) \right]  -A_{0} \tilde{H} -N_{j}H^{j}-\sigma P_{N}-\mu \pi,
\end{eqnarray}
here $\mathcal{V}=\mathcal{V}_{\text{g}}+\mathcal{V}_{\text{EM}}$. 

At this stage some comments are pertinent. In obtaining (\ref{Hamiltoniandensity}) we have fixed $\lambda$ to its critical point $1/3$ in Ho\v{r}ava's Hamiltonian. The consequence is that only the transverse-traceless tensorial modes and the transverse gauge potential propagate, they are the physical degrees of freedom of the theory \cite{BellorinRestucciaTello2018b,RestucciaTello2020,RestucciaTello2021}. The potential of the theory $\mathcal{V}(g_{ij},N,A_{i})$ contains then all scalar fields with high order derivatives, up to sixth order, of both sectors, the gravitational field, the vector $a_{i}$ and the gauge field. The action  (\ref{eq01}), is invariant under diffeomorphisms on the spacelike leaves of the foliation and under reparametrizations of the time variable. The infinitesimal generators of these symmetries are
\begin{equation}
    \delta x^{i}=\xi^{i}(t,x^{j}), \quad \delta t=f(t).
\end{equation}
Besides, the theory is invariant under $U(1)$ local gauge transformations. In fact, the infinitesimal transformations of the canonical variables $\{g_{ij}, A_{i}, N\}$ given by 

\begin{eqnarray} \label{eq29}
\delta g_{ij}&=& \partial_{i}\xi^{k}g_{jk}+\partial_{j}\xi^{k}g_{ik}+\xi^{k}\partial_{k}g_{ij}+f\dot{g}_{ij},  \\ \label{eq31}
\delta N&=& \xi^{k}\partial_{k}N+f\dot{N}+\dot{f}N,\\ \label{eq32}
\delta A_{i}&=&\partial_{i}\xi^{k}A_{k}+\xi^{k}\partial_{k}A_{i}+f\dot{A}_{i}+\partial_{i}\zeta,\\
\delta A_{0}&=&f\dot{A}_{0}+\dot{f}A_{0}+\xi^{k}\partial_{k}A_{0}+\dot{\xi}^{k}A_{k}+\dot{\zeta}
\end{eqnarray}
Note that the potential scale as $A_{i}\rightarrow b^{0}A_{i}$,  		$A_{0}\rightarrow b^{-z+1}A_{0}$.

show that the 3-dimensional metric transforms as a tensor and scalar field under spacelike diffeomorphisms and time reparametrizations, respectively. Whereas, the lapse function $N$ behaves as a scalar under spacelike diffeomorphisms and scalar density under time reparametrizations. Finally, the gauge vector-field $A_{i}$ transforms as a vector field under spacelike diffeomorphisms, a scalar field under time reparametrizations, and a gauge field under $U(1)$ gauge group. Regarding this point, the last term of (\ref{eq32}) represents the $U(1)$ gauge transformation of the vector $A_{i}$, whose infinitesimal generator is $\zeta$. Under gauge transformation both $g_{ij}$ and $N$ are invariant whilst $A_{i}$ transforms as a gauge field
\begin{equation}
    \delta_{\zeta} g_{ij}=\delta_{\zeta}N=0, \quad \delta_{\zeta}A_{i}=\partial_{i}\zeta.
\end{equation}
On the other hand, under the mentioned symmetry laws $A_{0}$, $N_{i}$, $\sigma$ and $\mu$ transform as Lagrange multipliers.

Next, we shall discuss the form of the full potential $\mathcal{V}(g_{ij},N,A_{i})$ of the theory. In the original Ho\v{r}ava's proposal, for pure anisotropic gravity, the complete potential of the theory up to $z=3$ derivatives contains around 100 terms. However, this long list, for pure anisotropic gravity, is greatly reduced if one is interested in considering only those objects  relevant to the propagator of the physical degrees of freedom which, indeed, are the only terms that contribute to the wave zone. Consequently, only those terms contributing to the wave zone will be taken into account. These terms are quadratic in the fields $R_{ij}$, $a_{i}$ and $F_{ij}$. So, the general form of the potential considering all possibilities up to $z=3$ is given by the sum of
\begin{align}\label{eq54}
{\mathcal{V}}^{(z=1)}=&-{{R}}+\frac{1}{4}F_{ij}F^{ij}-{\alpha a_{i}a^{i}}, \\ \label{eq55}
{\mathcal{V}}^{(z=2)}=&-\beta_{1}{R}_{ij}{R}^{ij}-\kappa_{1}\nabla_{l}F^{\ l}_{i}\nabla_{m}F^{im}-\beta_{2}{R}^{2}-{\alpha_{1}{R}\nabla_{i}a^{i}}-{\alpha_{2}\nabla^{i}a^{j}\nabla_{i}a_{j}}, \\ \label{eq56}
{\mathcal{V}}^{(z=3)}=&-\beta_{3}\nabla_{i}{R}_{jk}\nabla^{i}{R}^{jk}-\kappa_{2}\nabla_{k}\nabla_{l}F^{\ l}_{i}\nabla^{k}\nabla_{m}F^{im}-\beta_{4}\nabla_{i}{R}\nabla^{i}{R}-\alpha_{3}\nabla^{2}{R}\nabla_{i}a^{i} \nonumber \\ &-\alpha_{4}\nabla^{2}a_{i}\nabla^{2}a^{i}.
\end{align}	

At this stage, some comments are in order. First, all terms presented in (\ref{eq54})-(\ref{eq56}) are invariant under symmetries of the theory and so is the full potential of the theory. Secondly, in principle, all coupling constants for the gravitational and gauge sector could be different ($\beta's$ and $\kappa's$). However, taking into account both, theoretical arguments and recent experimental data, in the IR limit ($z=1$ terms), the propagation speed of the gravitational and electromagnetic waves are exactly the same or they differ at most in one part of $10^{-15}$. So, we keep at $z=1$ the same coupling constant $\beta=1$ for both sectors, since in this anisotropic non-relativistic theory, $\beta$ ends up being the propagation speed at low energies of both the gravitational and gauge vector field physical degrees of freedom. In this way, a comparison with the well-established Einstein-Maxwell theory can be easily performed.

It turns out that, at low energies where only the $z=1$ potential terms are relevant, for $\alpha=0$ and $\beta=1$, the field equations of both models, the one obtain from a dimensional reduction and the model we propose here, exactly agree with the Einstein-Maxwell theory in a particular gauge \cite{RestucciaTello2020}. 

Finally, we discuss the constraint structure of the theory. From the Hamiltonian (\ref{Hamiltoniandensity}), the theory posses four primary constraints $\tilde{H}$, $H^{j}$, $P_{N}$ and $\pi$. The former ones, $\tilde{H}$ and $H^{j}$, correspond to first-class constraints, generators of the $U(1)$ gauge symmetry, and the spacelike diffeomorphisms, respectively. Specifically, these constraints are given by   
\begin{eqnarray}
\label{H-tilde_constraint}
\tilde{H}&\equiv&\partial_{i}E^{i}=0, \\ 
\label{Hj_constraint}
H^{j}&\equiv& 2\nabla_{i}\pi^{ij}+E^{i}g^{jk}F_{ik}=0.
\end{eqnarray}
It should be pointed out, that constraint (\ref{H-tilde_constraint}) is equivalent to the Gauss law in the Maxwell theory. Concerning (\ref{Hj_constraint}), the first term in the right-hand member generates spacelike diffeomorphisms on the pair $\{g_{ij},\pi^{ij}\}$, while the second object generates the same symmetry on the gauge field $A_{i}$. This term is important, in order to guarantee the correct transformation law (\ref{eq31}) of the gauge vector field $A_{i}$ (and its conjugate momentum $E^{i}$). Additionally, this constraint (the so-called momentum constraint) can be supplemented by another extra piece $N\partial^{j}P_{N}$, generator of the spacelike diffeomorphisms on the lapse function $N$ and its conjugate momentum $P_{N}$ \cite{Bellorin:2011ff,Donnelly:2011df}. This can be done since $P_{N}$ vanishes on the phase space constrained surface.

The remaining constraints, $P_{N}$ and $\pi$ given by,
\begin{equation}
P_{N}=0, \quad \pi\equiv g_{ij}\pi^{ij}=0,
\end{equation}
are second-class constraints arising from the non-existence of time derivatives of $N$ and by the fact that $\lambda$ has been fixed to its critical point (the KC point $\lambda=1/3$), respectively. The conservation in time of these constraints leads to another second-class constraints
\begin{equation}
\label{HN_constraint}
\begin{split}
\dot{P}_{N}\approx 0 \Rightarrow H_{N}\equiv -\frac{1}{\sqrt{g}}\bigg[\pi^{ij}\pi_{ij}+\frac{E^{i}E_{i}}{2} \bigg]+\sqrt{g}\mathcal{U}=0,
\end{split}
\end{equation}
\begin{equation}
\label{HP_constraint}
\begin{split}
\dot{\pi}\approx0 \Rightarrow H_{\pi}\equiv\frac{N}{4\sqrt{g}}\bigg[6\pi^{lm}\pi_{lm}+E^{k}E_{k}\bigg]-\sqrt{g}\mathcal{W}=0.
\end{split}
\end{equation}
In the above expressions $\mathcal{U}$ and $\mathcal{W}$ correspond to 
\begin{equation}
\mathcal{U} \equiv \frac{1}{\sqrt{g}} \frac{\delta}{\delta N} \int d^{3} y \sqrt{g} N \mathcal{V}=\mathcal{V}+\frac{1}{N} \sum_{r=1}(-1)^{r} \nabla_{i_{1} \cdots i_{r}}\left(N \frac{\partial \mathcal{V}}{\partial\left(\nabla_{i_{r} \cdots i_{2}} a_{i_{1}}\right)}\right)
\end{equation}
and,
\begin{equation}
    \mathcal{W} \equiv g_{i j} \mathcal{W}^{i j}, \quad \mathcal{W}^{i j} \equiv \frac{1}{\sqrt{g} N} \frac{\delta}{\delta g_{i j}} \int d^{3} y \sqrt{g} N \mathcal{V},
\end{equation}
where $\nabla_{ij\ldots k}$ stands for $\nabla_{i}\nabla_{j}\ldots \nabla_{k}$.


\subsection{Wave zone and  stability of the model }

In this section, we analyze the stability of the Minkowski metric as the background on which the gravity and electromagnetic waves propagate. We notice that the Minkowski metric is a solution of the field equations of the model (\ref{eq01}). In fact, it is a solution with enhanced symmetries compared to the anisotropic formulation. In order to study the  stability of the background we consider the quadratic Hamiltonian arising from (\ref{eq01}). The stability requirement reduces to show that the elliptic operators $\beta \nabla + \beta_{1} \nabla^{2} - \beta^ {3} \nabla^{3}$ and $\beta \nabla + 2 \kappa_{1}\nabla^{2} + 2 \kappa^{2} \nabla^ {3}$ are strictly positive definitive. The stability conditions for $2 \kappa_{1} $ and $2 \kappa_{2}$ are given below in (\ref{DesigualdadKappa3Negativo}) and (\ref{desigualdadKappa1Cuadrado})  and similar for $\beta_{1}$ and $\beta_{3}$ respectively. Under such assumptions, we can now analyze the existence of a wave zone.
It is known that in both Einstein's General Relativity and in the Ho\v{r}ava-Lifshitz gravity theory, there exists a well-defined wave zone for asymptotically flat solutions \cite{ArnowittDesserMisner1961,MestraPenaRestuccia2021, MestraPenaRestuccia2021a}. In the wave zone, the dominant mode $\mathcal{O}(1/r)$ of the $g^{TT}_{ij}$ component, the transverse-traceless tensorial modes, satisfies a linear equation with constant coefficients. Although there exists  a non-trivial Newtonian background of the same order as the dominant mode, it does not interfere with the propagation of the $TT$ modes.

In the low energy case, the modes satisfy a linear relativistic ($\beta=1$) wave equation
\begin{equation}
\label{WaveEquationForgTT}
\ddot{g}^{i j T T}-\beta \Delta g^{i j T T}=0,
\end{equation}
and, at high energies, these modes satisfy a generalized anisotropic wave equation
 \begin{equation}
     	\label{wave-six-gpuntoTT}
	\ddot{g}_{ij}^{TT}-\left(\beta \Delta+\beta_{1}\Delta^{2}-\beta_{3}\Delta^{3}\right)g_{ij}^{TT}=0,
 \end{equation}
where high-order derivative operators arise in a natural way in the Ho\v{r}ava-Lifshitz scenario.

These equations have a solution of the form 
\begin{equation}
    g_{ij}^{TT} \sim \frac{e^{-i\omega t+ ikr}}{r},
\end{equation}
where the dispersion relation, for the wave equation (\ref{wave-six-gpuntoTT}), obtained when high order derivative operators are considered, is \cite{MestraPenaRestuccia2021a} 
\begin{equation}
   \omega^{2}(k)=\beta k^{2}-\beta_{1}k^{4}-\beta_{3} k^{6}. 
\end{equation}

In \cite{BellorinRestucciaTello2018b,RestucciaTello2020}, a 3+1 dimensional anisotropic gravity model coupled to a gauge field was proposed. This model is obtained from 4+1 dimensional Ho\v{r}ava's gravity via a dimensional reduction. It was shown in \cite{Mestra-Restuccia-Tello2022}, that a well-defined wave zone at low energies, exists for this model. Under the same hypothesis,  it is possible to show  for the particular model considered in this work, that a well-defined wave zone exits when all high-order derivative operators for both, the gravitational and electromagnetic sectors, are taken into account. The wave zone again has a dominant order $\mathcal{O}(1/r)$ and, both the $TT$ modes for the gravitational field and the $T$ modes for electromagnetism interaction satisfied a linear wave equation. We will impose the  requirements given in \cite{Mestra-Restuccia-Tello2022}, section (IIIA), on the wave zone to determine the behavior at high energy of the physical degrees of freedom. Given these requirements, we solve the field equations and show  that there exists a zone where these requirements are satisfied and we determine then the evolution equations for the physical degrees of freedom: the wave equations. We notice from the interaction terms of the potential (\ref{eq54})-(\ref{eq56}), the only ones that contribute to the wave zone, that there is no interaction term that couples the electromagnetic $F_{ij}$ to the pure gravity spacelike vector $a_{i}$ and the Ricci tensor $R_{ij}$ of the spacelike leaves of the foliation. Although there is a gravity-electromagnetic coupling through the metric, this property implies that at the leading order both wave equations are decoupled.  Therefore, in the wave zone, the physical degrees of freedom of the gravitational field do not interact with the physical degrees of freedom of the electromagnetic field, furthermore, none  of them interact with the Newtonian background whose order again is $\mathcal{O}(1/r)$. This is analogous to what happens in GR coupled to Maxwell's theory \cite{ArnowittDesserMisner1961, ArnowitDesrMisnert2008}.
Besides, the energy and momentum of the system are given by the contribution of both sector \i.e., the gravitational and electromagnetic interactions.

Taking into account the previous comment on the potential interaction terms, it follows that the estimates in orders of $1/r$ follow by similar arguments as in \cite{MestraPenaRestuccia2021}. So,
from the constraints (\ref{Hj_constraint}) and $\pi=0$ we obtain
\begin{equation}
\begin{split}
 \label{EstimatePiLongitudinal}
    \pi^{Lij}\sim \pi^{Tij}\lesssim \frac{\tilde{B}}{r^{2}}+k\frac{\tilde{A} e^{ikr}}{r^{2}}\,. 
\end{split}
\end{equation}

Using the gauge condition  $g_{ij,j}=0$, the constraints (\ref{HP_constraint}) and (\ref{HN_constraint}) imply 
 \begin{equation}\label{napporx}
	N-1 \sim g^{T} \lesssim\frac{B}{r} +\frac{\hat{A}  e^{ikr}}{r^{2}},
	\end{equation} 
we notice that they are of the order of $\mathcal{O}(1/r)$, but only in the non-oscillatory part. This is the main reason why they do not contribute to the equations describing the propagation of the physical degrees of freedom.

Finally, from  equations of motion at order $\mathcal{O}(1/r)$ for the gravity degrees of freedom we obtain the behavior of $\pi^{ijTT}$, which  is oscillatory of order $\mathcal{O}(1/r)$ and the wave equation for the TT part of the metric
 \begin{eqnarray}
\ddot{g}_{ij}^{TT}=(\beta\Delta+ \beta_{1}\Delta^{2}-\beta_{3}\Delta^{3})g_{ij}^{TT}.
\end{eqnarray}
{For the electromagnetic sector, the constraint (\ref{H-tilde_constraint}), implies that $A_{i}$ is a transverse mode. In the wave zone the equations of motion at dominant order $\mathcal{O}(1/r)$ for the gauge field reduce to}
\begin{eqnarray}
\label{GeneralizedWaveEquation}
\ddot{A}_{i}^{T}=\left(\beta \Delta +2 \kappa_{1}\Delta^{2}-2\kappa_{2}\Delta^{3}\right)A^{iT}.
 \end{eqnarray}
Both the gravitational and electromagnetic degrees of freedom have a spherical wave solution of the form  
\begin{equation}
    A_{i}^{T} \sim \frac{e^{-i\omega t+ ikr}}{r},
\end{equation}
 with dispersion relation (for the gauge potential $A^{T}_{i}$, and analogous for $g^{TT}_{ij}$)
\begin{equation}
    \label{RelacionDispersiónElectromagnetica}
   \omega^{2}(k)=\beta k^{2}-\hat{\kappa}_{1}k^{4}-\hat{\kappa}_{3} k^{6}, 
\end{equation} 
{where $\hat{\kappa}_1=2\kappa_{1}$ and $\hat{\kappa}_{2}=2\kappa_2$. This solution represents dispersive waves whose phase velocity, $v_{f}\equiv \omega/k$, is given by the relation
\begin{equation}
 v_{f}^{2}=\beta-\hat{\kappa}_{1}k^{2}-\hat{\kappa}_{3}k^{4}.  
\end{equation}
To guarantee the square phase velocity positivity and thus the stability of the solutions, we require 
\begin{eqnarray}
\label{DesigualdadKappa3Negativo}
        \hat{\kappa}_{3}<0,
 \end{eqnarray}       
        together with a) $\hat{\kappa}_{1}\leq 0$ or b) $\hat{\kappa}_{1}>0$ and \begin{eqnarray}\label{desigualdadKappa1Cuadrado}
        \hat{\kappa}_{1}^{2}<4\beta \left|\hat{\kappa}_{3}\right|.
\end{eqnarray}
It should be noted that phase velocity in the vacuum is a function of $k$ in contrast with its relativistic counterpart which is constant.}

The dispersion relation may be rewritten in terms of the energy and momentum as

\begin{equation}
  E(p)^{2}={\beta} p^2- \hat{\kappa}_{1}p^{4}-\hat{\kappa}_{3}p^{6}.
\end{equation}
The group velocity $v \equiv \partial \omega/\partial k$ becomes then
\begin{equation}
v(p)=\frac{\beta p-2\hat{\kappa}_{1} p^{3}-3\hat{\kappa}_{3} p^{5}}{\sqrt{\beta p^{2}-\hat{\kappa}_{1} p^{4}-\hat{\kappa}_{3} p^ {6}}}.
\end{equation}

From (\ref{DesigualdadKappa3Negativo}) the fifth power term  on the momentum is positive, hence if $\hat{\kappa}_{1}<0$ the group velocity  is $v>\sqrt{\beta}$ while if $\hat{\kappa}_{1}>0$ the group velocity is $v<\sqrt{\beta}$ if the third power on the momentum term is much bigger than the fifth power one. In what follows we will take $\sqrt{\beta}=c=1$.

\subsection{The dispersion relation}

Recently LHAASO reported the detection of more than 5000 very high energy photons from gamma-ray burst GRB 221009A with energies above 500 GeV up to 18 TeV \cite{li2022lorentz}. Very high energy photons $\gamma$ can interact with background photons $\gamma_{b}$, such as those from the cosmic microwave background (CMB) and the extragalactic background light (EBL), and produce an electron-positron pair
$\gamma \gamma_{e}\rightarrow e^{-}e^{+}$. According to relativistic physics, there is a threshold for this interaction 
\begin{equation}
\label{ThrestoldRelativistic}
    E>E_{\text{th}}=\frac{m_{e}^{2}}{\epsilon_{\text{b}}},
\end{equation}
where $E$ is the energy of $\gamma$ and $E_{b}$ the energy of $\gamma_{b}$, $m_{e}$ is the electron mass. For photons with $E>E_{\text{th}}$ the gamma-ray is strongly attenuated. For CMB photons $E_{\text{th}}\cong 411$ TeV. Since the detected VHE photons have as maximum energy 18 TeV, the CMB background is transparent to the $\gamma$ photons. However, this is not the case for the EBL for which the threshold energy is 261 GeV to 261 TeV. So, according to relativistic physics, most of the photons should have been attenuated. If the high-energy photons travel extragalactic distances, it has been argued that they should not satisfy a relativistic dispersion relation. A MDR with subluminal photon velocities has been proposed as an explanation for the observation of these high energy photons \cite{LiMa2021Ultra}. The scale energy of the breakdown of the Lorentz symmetry according to a MDR
\begin{equation}\label{A}
    \omega^{2}=k^{2}\left(1-\frac{\xi_{n}}{n}k^{n}\right), \quad (\text{we use }\, c=1),
\end{equation}
has been estimated from Fermi laboratory data using the difference in arrival time, of photons of different energies emitted from the same source. The accepted characteristic parameter of the MDR, for $n=1$ is  $\xi_{1}=(E_{\text{LV}_{,1}})^{-1}$, $E_{\text{LV}_{,1}}=3.6 \times 10^{17}$ GeV \cite{xu2018regularity}, and for $n=2$ $\xi_{2}=(E_{\text{LV}_{,2}})^{-2}$, $E_{\text{LV}_{,2}}=6.8 \times 10^{9}$ GeV \cite{zhang2015lorentz}.
Under the same assumption used in  \cite{li2021threshold}, that is the energy-momentum conservation of the $\gamma \gamma_{b}\rightarrow e^{-}e^{+}$
interaction, one obtains the general relation  
\begin{equation}
\label{B}
    m^{2}_{e}=\left(\frac{\omega+\epsilon_{\text{b}}}{2}\right)^{2}-\left(\frac{k-\epsilon_{\text{b}}}{2}\right)^{2}.
\end{equation}

We will now replace the dispersion relation (\ref{RelacionDispersiónElectromagnetica}) for photons arising from the model (\ref{eq01}). We redefine the coupling parameters in order to have a more direct comparison with the MDR in (\ref{A}). We have 
\begin{equation}\label{C}
   \omega^{2}=k^{2}\left(1-\xi k^{2}+\frac{\xi^{2}}{a^{2}}k^{4}\right), 
\end{equation}
where $a$ is a new parameter. We notice that this MDR is a closed relation, it is not an expansion into higher powers of $(E/E_{\text{LV}_{,n}})^{n}$ which are suppressed at energies well below $E_{\text{LV}_{,n}}$.

We are going to consider $\xi>0$, hence the positivity of the quadratic Hamiltonian, or equivalently the stability requirements imposes $\xi^{2}<\xi^{2}/a^{2}$. Consequently, we must have $a^{2}<1$, it is the only requirement on $a^{2}$. We denote

\begin{equation}
\label{DefXiTilde}
    \tilde{\xi}(k)\equiv \xi\left(1-\frac{\xi k^{2}}{4a^{2}}\right)
\end{equation}
and assume $k^{2}|\tilde{\xi}(k)|<<1$, which can be checked to be valid in the following discussion.

\begin{figure}[H]
  \centering
  \includegraphics[width=0.42\textwidth]{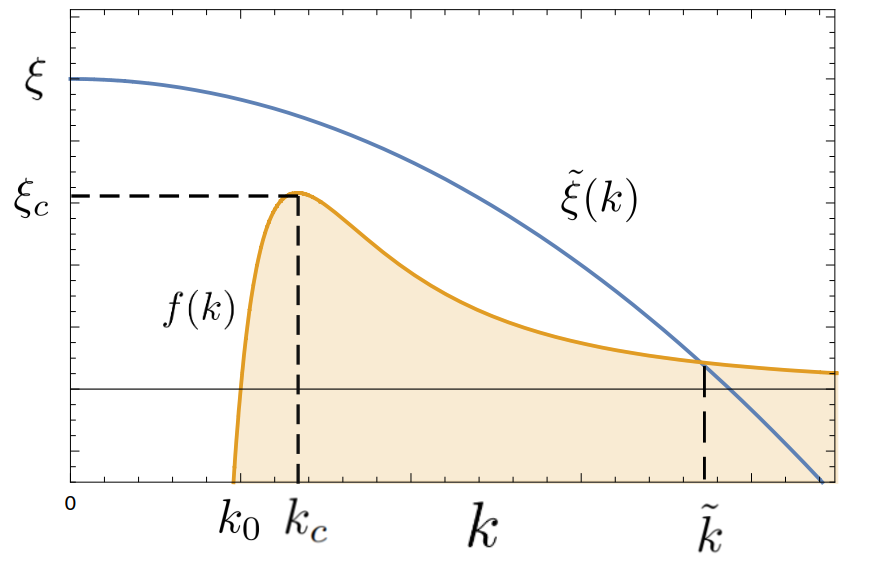}   \caption{Shows both functions $f(k)$ and $\tilde{\xi}(k)$, the colored zone represents the region where the photons are strongly attenuated. }\label{Graf}
\end{figure}

After replacing (\ref{C}) in (\ref{B}), we end up with 
\begin{equation}
    \tilde{\xi}(k)=\frac{4}{k^{4}}\left(\epsilon_{\text{b}}k-m^{2}_{e}\right),
\end{equation}
where $f(k)\equiv \frac{4}{k^{4}}\left(\epsilon_{\text{b}}k-m^{2}_{e}\right)$ is the same function which appears in the analysis of the $\gamma \gamma_{b}\rightarrow e^{-}e^{+}$ interaction using the MDR (\ref{A}) with $n=2$ \cite{li2021threshold}.  In this case the function $\tilde{\xi}(k)$ reduces to a constant $\xi$. For a given value of $\xi$ the intersection with $f(k)$ defines two values of $k$. For any $k$ between these two values, the photon $\gamma$ is strongly attenuated. For $k$ outside this set the background becomes transparent. In the FIG. \ref{Graf}  we show both functions $f(k)$ and   $\tilde{\xi}(k)$, defined in (\ref{DefXiTilde}).

The LIV model, when $\xi>\xi_{c}$, ensures that the background photons are transparent to the high energy photons, this is the case for MDR with $n=1$. However, the anisotropic model (\ref{eq01}) ensures that there is always (also when $\xi\leq \xi_{c}$) an attenuation due to the interaction $\gamma \gamma_{b}\rightarrow e^{-}e^{+}$, in the case of figure \ref{Graf}  for all $k>\tilde{k}$.

\begin{figure}[H]
  \centering
  \includegraphics[width=0.42\textwidth]{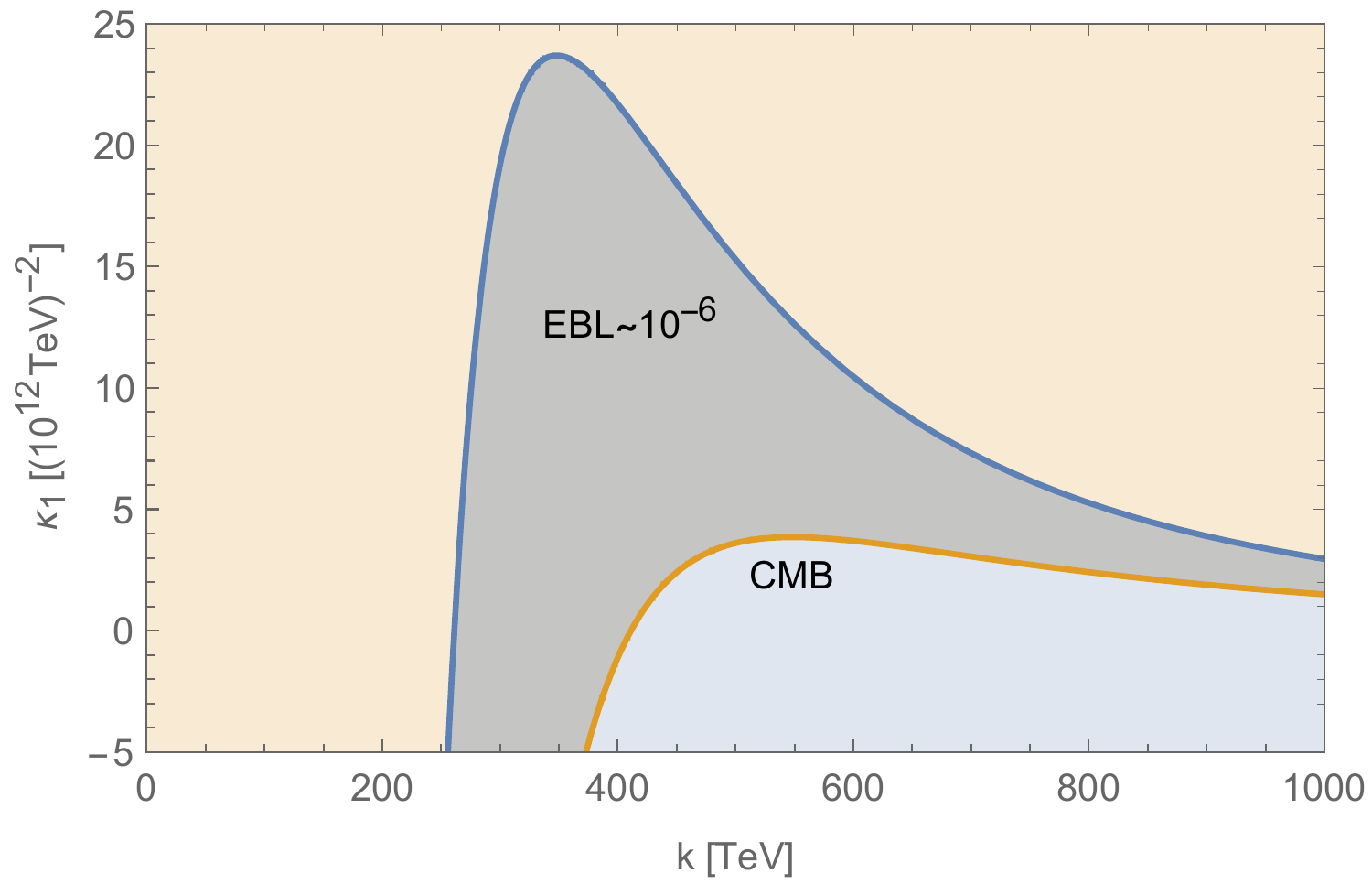} \,
  \includegraphics[width=0.45\textwidth]{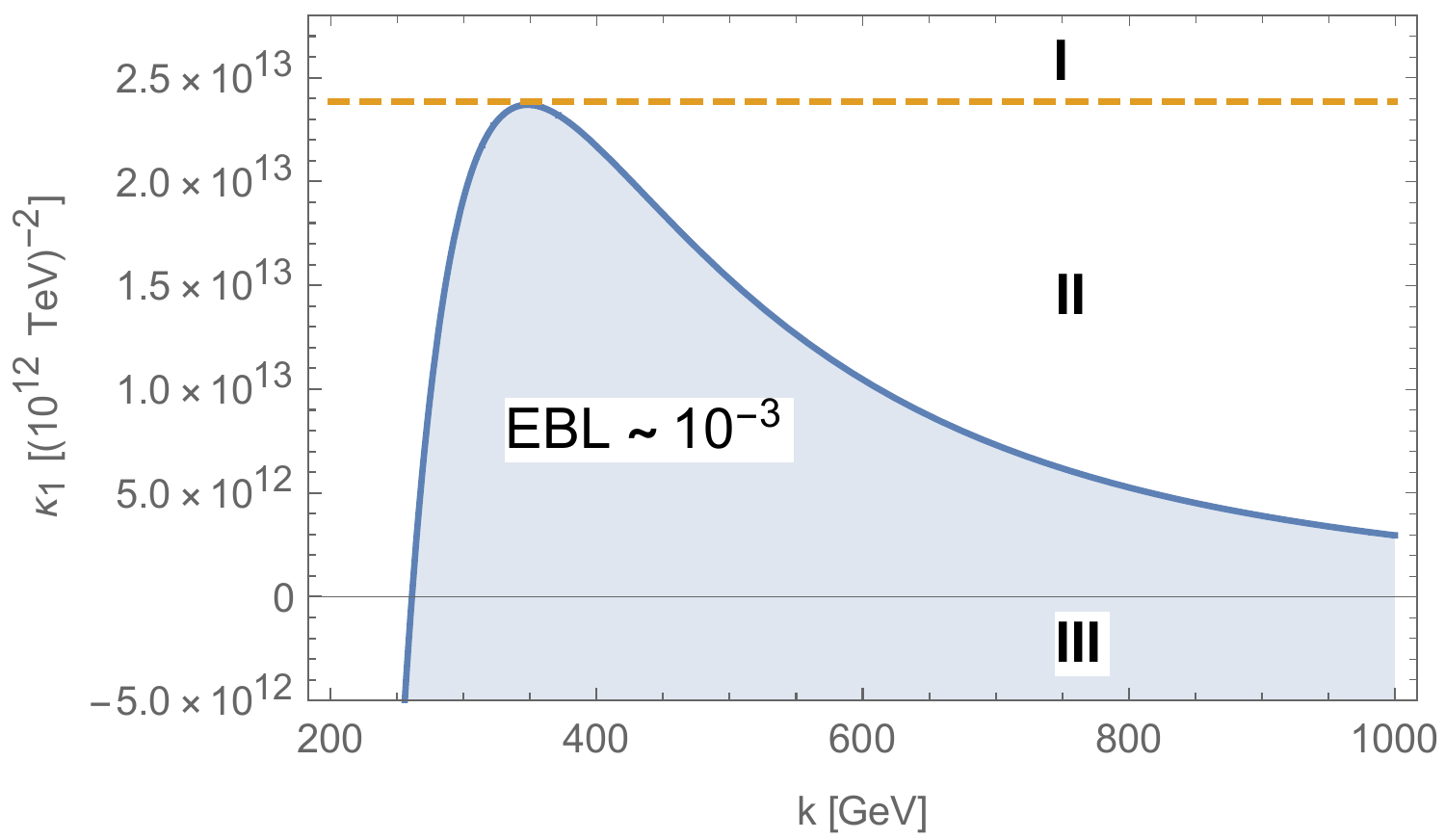}
  \caption{shows the function $f(k)$ for the MDR (\ref{C}) in the assumption that the term of highest power of momentum, $k^{6}$, is negligible with respect to the $k^{4}$ term. For the EBL background, we consider two cases, for the values of $\varepsilon_{\text{b}}$ indicated in the figure. For CMB the relativistic threshold is $411 \text{ TeV}$. The critical value of $k$ at which $f(k)$ has its maximum value is $k_{c}= 548 \text{ TeV}$ and the maximum is $\xi_{c}= 3.85\times 10^{-48} 1/(\text{eV})^{2}$, the value of $\xi$ is $1/(E_{\text{LV}_{,2}})^{2} = 1/ 6.8 \times 10^{18} \text{eV}$. 
Then $\xi > \xi_{c}$, there is no attenuation from the background for any value of $k$. For the EBL background, the relativistic threshold varies between $261 \text{ GeV}$ to $261 \text{ TeV}$, the critical value $k_{c}$ becomes  $348 \text{ GeV} < k_{c} < 348 \text{ TeV}$ and the maximum of $f(k)$, $\xi_{c}$,  varies from 
$2.3\times 10^{-32}$  to  $2.3\times 10^{-43} 1/(\text{eV})^{2}$.
 The estimated value of $\xi$ is $2.1\times 10^{-38} 1/(\text{eV})^{2}$, it belongs to the region 2 as defined in \cite{li2021threshold}.  Hence $\xi < \xi_{c}$, and there may be attenuation from the background depending on the values of $k$.
 }\label{fig2}
\end{figure}

There is then a qualitative difference between both models. The MDR (\ref{C}) ensures the existence of a threshold in all cases, it depends on $k$ and $n$  the parameters $a$ and $\xi$. For the CMB the estimated values are $\epsilon_{\text{b}}=6.35\times 10^{-4}$ eV and of $k_{0}=411$ TeV, $k_{0}$ is the relativistic threshold. For the MDR (\ref{A}) with $n=1$: $\xi_{1}=E^{-1}_{\text{LV}_{,1}}$, $E_{\text{LV}_{,1}}=3.6\times 10^{17}$ GeV and the critical $\xi^{-1}_{c}=4.5\times 10^{23}$ GeV. Hence $\xi>\xi_{c}$ and there is no attenuation of the gamma-ray for any value of $k$, the high energy photons cannot be absorbed by photons with energy $\epsilon_{\text{b}}$. For $n=1$ both backgrounds the CMB and EBL are transparent to the propagation of very high energy photons. For the MDR (\ref{C}), considering $\frac{\xi^{2}k^{2}}{4a^{2}}<<1$, the MDR reduces to MDR (\ref{A}) with $n=2$. In this case, the estimated value of $\xi_{2}\cong 10^{-39} \frac{1}{(\text{eV})^{2}}$ and the critical values $\xi_{c}=\left(\frac{3}{4}\right)^{3}\frac{\epsilon_{\text{b}}}{k^{3}_{0}}\cong 10^{-48} \frac{1}{(\text{eV})^{2}}$, hence we also have $\xi>\xi_{c}$. The difference is that in this case although we have approximated the calculations by a MDR (\ref{A}) with $n=2$, there exists a threshold beyond which the highly energetic photons will be absorbed by the background. This bound is, however, much bigger than the energies of the detected $\gamma$ photons. The MDR (\ref{C}) is then compatible with the recently detected photons with very high energy, with respect to the CMB background. In the case of the EBL background, the estimates are not as precise as in the CMB. According to the reported data, for $n=2$, $\xi<\xi_{c}$, and in this case it has to be determined if the energies of the detected photons are contained on the attenuation range or not. More information is needed to determine the compatibility of the MDR  with the EBL background. We illustrate this situation in figure \ref{fig2}.

\section{Conclusions}
We considered an anisotropic model describing the interaction of gravity and electromagnetic forces in the context of Ho\v{r}ava-Lifshitz framework. We showed the consistency of the formulation, the stability of the Minkowski space-time solution, and the existence of a wave zone where both interactions propagate satisfying independent wave equations in the presence of a nontrivial Newtonian background. Finally, we analyzed the propagation of highly energetic photons, which in this anisotropic model satisfy a modified dispersion relation compared to the relativistic one. We used the data from Fermi-LAT, Fermi-GBM and LHAASO on the recently observed highly energetic gamma-ray bursts.

\section*{ACKNOWLEDGEMENTS}
 J. Mestra-P\'aez acknowledge financial support from  Beca Doctorado Nacional 2019 CONICYT, Chile.   N° BECA: 21191442. J. Mestra-P\'aez acknowledges the PhD program Doctorado en F\'isica menci\'on en F\'isica Matem\'atica de la Universidad de Antofagasta for continuous support and encouragement. F. Tello-Ortiz thanks the financial support by project ANT-2156 at the Universidad de Antofagasta, Chile.

\bibliographystyle{elsarticle-num}
\bibliography{Bib_Horava_Maxwell}

\end{document}